# A Taxonomy of Collaboration in Online Information Seeking

Gene Golovchinsky, Jeremy Pickens, and Maribeth Back
FX Palo Alto Laboratory, Inc.
3400 Hillview Ave, Bldg 4,
Palo Alto, CA 94304
{gene, jeremy, back}@fxpal.com

**ABSTRACT**
People can help other people find information in networked information seeking environments. Recently, many such systems and algorithms have proliferated in industry and in academia. Unfortunately, it is difficult to compare the systems in meaningful ways because they often define collaboration in different ways. In this paper, we propose a model of possible kinds of collaboration, and illustrate it with examples from literature. The model contains four dimensions: intent, depth, concurrency and location. This model can be used to classify existing systems and to suggest possible opportunities for design in this space.

**ACM Classification Keywords**
H.3 INFORMATION STORAGE AND RETRIEVAL: H.3.5 Online Information Services: *Data sharing*

**INTRODUCTION**
Although commonly conceived as a solitary activity, information seeking often involves collaboration with others [2][5][9]. Collaborative filtering or recommendation systems (e.g., [4]) are common examples of online collaborative search: the behavior of people – paths taken, documents seen, etc. – while looking for information is used to inform the behavior of others searching for similar information later. The people involved do not need to be aware of each other; the system generates appropriate information as dictated by the users' actions. Other approaches (e.g., [1][10]) assume co-located users who communicate with each other normally, construct queries, and share search results produced by the system.

In this paper, we propose a taxonomy of collaboration in support of information seeking designed to distinguish the various forms of online collaboration. The taxonomy consists of four dimensions (intent, depth, concurrency and location) that can be used to characterize various aspects of collaboration.

Our model is related to that proposed by Hansen and Järvelin [3]: we share the synchronous/asynchronous dimension and we are concerned with computer-mediated communication (although we do not rule out additional human-to-human communication channels). We differ somewhat in our treatment of coupling: CSCW literature uses the terms "loosely coupled" and "tightly coupled" to refer to a variety of phenomena, including organizational structure, software architecture, and degree of collaboration, among others. We focus on technical (rather than on social) issues in this model, and represent some notions of coupling in the "depth of mediation" and "explicit vs. implicit" dimensions.

In the rest of the paper, we first describe the dimensions on the model, and then illustrate them with examples from commercial systems and academic research.

**DIMENSIONS OF COLLABORATION**

**Intent: Explicit vs. implicit**
Recommender systems use behavior of a group of people who have engaged with particular content to suggest choices to others searching for similar information. The goal here is to use information previously found by others to inform new search results. This is implicit collaboration: while people may be generally aware that their results are based in part on data obtained from other users, they may not know who those people are or what purpose they had in mind while searching. Thus the collaboration here exists because the search engine uses historical data as a source of evidence for document relevance to a query. In some sense this is not strictly collaboration, but rather a coordination of people's activities.

We can contrast this with explicit collaboration, in which a small group of people searches for documents to meet a shared information need. The need may evolve over time, but through-out a search session that need is shared by all team members and it motivates their search activities. This is related to Morris and Teevan's notion of 'task based" collaboration [7].

**Depth of mediation**
Mediation of information seeking can occur in the user interface [1][6][10], and may also be reflected in the underlying search algorithms. We distinguish UI-only mediation from deeper algorithmic mediation that explicitly represents contributions from different people in the algorithms that retrieve information. Examples of this mediation include recommender systems (that use records of individuals' selections to rank documents for retrieval) and FXPAL's Cerchiamo [8] that combine relevance feedback from multiple people to rank documents and offer search term suggestions based on team members' actions.

**Concurrency: Synchronous vs. asynchronous**
This dimension reflects the flow of influence among members of a collaborating group. If search activity by more than one person occurs at the same time, it is possible for influence (see the Depth of mediation dimension below) to flow between members during a search session. The asynchronous case describes the condition in which people do not work at the same time; those who search later can benefit from the work of earlier collaborators, but the earlier ones did not benefit from contributions of subsequent collaborators.

This does not mean that team members engaged in synchronous collaboration need to operate in lock-step, searching or browsing results simultaneously. Rather it means that they are actively involved in various aspects of information seeking activity *at the same time*. They may divide their activities in any manner supported by the tools they use; the key is the possibility that each team member's actions can influence other team members.

**Location: Co-located vs. distributed**
Finally, collaboration may be co-located or distributed. Distributed collaboration implies the need for additional channels to coordinate searchers' activities. Such channels may include chat, voice, or audio conferencing.

## EXAMPLES OF COLLABORATIVE SYSTEMS
In the previous section we introduced the four dimensions of our taxonomy. We now introduce a prioritization or ordering of what we believe to be the most important dimensions for distinguishing between existing and future online, collaborative retrieval systems. Most significant is searcher intent. Collaborative retrieval systems built for users with an explicitly shared information need will most likely be very different from systems built to support users with similar behavioral patterns, but who do not necessarily approach their retrieval activity with the same information need. Depth of mediation and concurrency are the next most significant dimensions and location, while important, plays a tertiary role after intent. Therefore, in the remainder of this section, we will provide examples of collaborative information retrieval systems that have been factored by the intent dimension.

**Examples of Implicit Collaboration**
Some familiar Web 2.0 collaborative systems such as Amazon shopping recommendations [4], Google Personalized search,[1] and iSpy [11] are good examples of implicit, asynchronous, deeply-mediated distributed search. The systems obviously differ in the details of their algorithms, but they also differ in the scope of the group whose behaviors influence search results. Amazon pools everyone's behaviors, Google's personalized search attempts to infer group membership from latent factors, and iSpy establishes membership based on organizational affiliation. Collaboration is implicit in all cases because each person's search intent is unknown to all others (despite varying degrees of user control over group membership), whether they are searching on similar topics or not. Collaboration is deeply mediated because each person's inputs affect the group's search results, and of course their interactions are asynchronous and distributed.

**Examples of Explicit Collaboration**
Several recent systems have been described in the literature that are designed to support explicit, synchronous collaboration [5][6][8][10]. These systems are designed to allow small teams to work together to search for relevant information. Some (e.g., SearchTogether [6]) can work in co-located or distributed fashion, while others have been designed to support co-located search. DiamondTouch Fischlar [10], SearchTogether [6] and CoSearch [1] mediate collaboration in the interface only: the systems allow users to compose queries – either singly or jointly – and then execute these queries one a time. Search results are then displayed to the users (or optionally partitioned among users in SearchTogether).

Cerchiamo [8], on the other hand, implements a deeper form of mediation: while each team member searches and browses results independently, the system coordinates their judgments of relevance, and offers search term suggestions based on team partners' actions. Furthermore, the two team members act in different roles – Prospector to discover potentially relevant documents, and Miner to explore such groupings – and therefore use different interfaces. The system mediates their activities, enabling the team to discover more, and different, relevant documents than they would by working separately in parallel.

## CONCLUSION
We have proposed a four-dimensional model for collaborative search that can be used to characterize existing systems. The dimensions of the model – intent, depth, concurrency and location – can be used to classify existing systems and may also be used to predict interesting points in the design space: what would an explicit asynchronous system look like? What useful tasks would a synchronous implicit system solve?

---

[1] www.google.com/history